\documentclass{PoS}
\usepackage{bbm, bm, epsfig, picins, amsmath, amsfonts, subfigure, amssymb, fontenc, times, mathptmx, graphicx}

\title{The imbalanced Fermi gas at unitarity}

\ShortTitle{The imbalanced Fermi gas at unitarity}

\author{\speaker{O. Goulko}\\
        DAMTP, University of Cambridge, Wilberforce Road, Cambridge CB3 0WA, UK\\
        E-mail: \email{O.Goulko@damtp.cam.ac.uk}}

\author{M. Wingate\\
        DAMTP, University of Cambridge, Wilberforce Road, Cambridge CB3 0WA, UK\\
        E-mail: \email{M.Wingate@damtp.cam.ac.uk}}

\abstract{Lattice field theory is a useful tool for studying strongly interacting theories in condensed matter physics. A prominent example is the unitary Fermi gas: a two-component system of fermions interacting with divergent scattering length. With Monte Carlo methods this system can be studied from first principles. In the presence of an imbalance (unequal number of particles in the two components) a sign problem arises, which makes conventional algorithms inapplicable. We will show how to apply reweighting techniques to generalise the recently developed worm algorithm to the imbalanced case, and present results for the critical temperature, the energy per particle, the chemical potential and the contact density for equal, as well as unequal number of fermions in the two spin components.}

\FullConference{The XXVIII International Symposium on Lattice Filed Theory\\
                 June 14-19,2010\\
                 Villasimius, Sardinia Italy}

\begin{document}

\relpenalty=9999
\binoppenalty=9999
\widowpenalty=0
\clubpenalty=0

\renewcommand{\Im}{\textnormal{Im}}
\newcommand{\sgn}{\textnormal{sign}}
\newcommand{\Tr}{\textnormal{Tr}}

\section{Introduction}
With the discovery of the renormalization group came the birth of both lattice gauge theory and statistical field theory as we know them today. Nowhere is this kinship more obvious than in the study of critical phenomena. For example, the numerical studies of deconfinement in gauge theories are identical in most respects to those of demagnetization in spin systems. Now a challenge in both fields is to develop more efficient methods for systems with strongly interacting fermions. Then we could better understand the phases of gauge theories with matter and of superfluids and superconductors.

Given that the common problem lies in dealing with fermionic degrees of freedom, we are motivated to consider the unitary Fermi gas, which is beautiful in its simplicity. This is a system of 2-component nonrelativistic fermions (typically ultracold ${}^{6}$Li or ${}^{40}$K atoms in experiments) interacting via a short-range potential. The gas is so dilute that only s-wave scattering is relevant, thus it is sufficient to treat the interaction as a local 4-fermion coupling. On a spatial lattice, the simplest Hamiltonian is that of the attractive Hubbard model.

The lattice theory describing unitary Fermi gases avoids some of the complications of lattice gauge theory. With nonrelativistic fermions there is no chiral symmetry, so there is no fermion doubling problem. Without nonabelian gauge fields, there are no topological sectors to worry about sampling with the correct distribution. This is an ideal system for concentrating on Monte Carlo methods for fermions. In addition to both being strongly coupled theories, QCD and unitary Fermi gas both exhibit spontaneous symmetry breaking at low temperatures, restored at some critical temperature. Like finite density QCD, the fermion matrix of the unitary gas can have a nonpositive determinant.

\vspace{-2mm}
\section{Setup}
We start from the Fermi-Hubbard model, which in the grand canonical ensemble reads,
\begin{equation}
H=H_0+H_1=\sum_{\mathbf{k},\sigma}(\epsilon_\mathbf{k}-\mu_\sigma)c^\dagger_{\mathbf{k}\sigma}c_{\mathbf{k}\sigma}+U\sum_{\mathbf{x}}c^\dagger_{\mathbf{x}\uparrow}c_{\mathbf{x}\uparrow}c^\dagger_{\mathbf{x}\downarrow}c_{\mathbf{x}\downarrow},
\end{equation}
where $\epsilon_\mathbf{k}=\frac{1}{m}\sum_{j=1}^{3}(1-\cos{k_j})$ is the discrete dispersion relation, $\mu_\sigma$ the chemical potential and $c^\dagger_{\mathbf{k}\sigma}$ ($c_{\mathbf{k}\sigma}$) the time-dependent fermionic creation (annihilation) operator. We use $\hbar=k_B=2m=1$. The index $\sigma\in\{\uparrow,\downarrow\}$ labels the fermionic species. The coupling constant $U<0$ corresponding to attractive interaction can be tuned so that the scattering length becomes infinite. The corresponding value is $U=-7.914$. We work on a 3D periodic lattice with $L^3$ sites. The continuum limit can be taken by extrapolation to vanishing filling factor $\nu=\langle \sum_{\sigma}c^\dagger_{\mathbf{x}\sigma}c_{\mathbf{x}\sigma}\rangle\rightarrow0$.

According to \cite{rubtsov} the partition function for this model can be written as a series of products of two matrix determinants built of free finite-temperature Green's functions. If $\mu_\uparrow=\mu_\downarrow$ (the balanced case) these determinants are equal so that all terms in the series are positive and it can be used a probability distribution for Monte Carlo sampling. The order parameter for the phase transition is given by a two-point correlation function of the operator $c_{\mathbf{x}\uparrow}c_{\mathbf{x}\downarrow}$. To obtain $T_c$ from the numerical data, previous work \cite{burovski,bulgac1} used a procedure involving an approximation which introduced a systematic error. We have improved the data analysis method so that this approximation is no longer necessary \cite{ourmain}. Our final result will be the dimensionless quantity $T_c/\varepsilon_F$, where the Fermi energy $\varepsilon_F=(3\pi^2\nu)^{2/3}$ is the only energy scale of the system. The continuum limit is taken by obtaining $T_c/\varepsilon_F$ for different values of $\nu$ and extrapolating to vanishing filling factor \cite{burovski, chenkaplan}.
\begin{figure}
\vspace{-4mm}
\begin{center}
\includegraphics[width=0.45\textwidth]{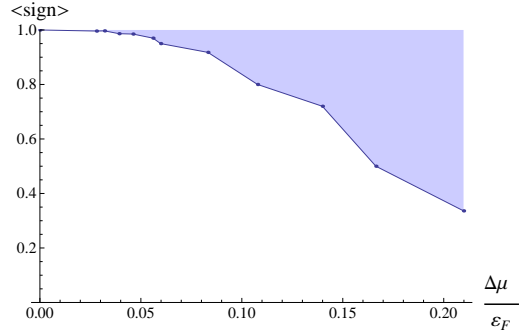}
\caption{Schematic plot of the average sign near the critical point. The shaded area covers the range of values the sign can take at different values of lattice size and chemical potential. The lower boundary of this area is the ``worst-case'' curve of the sign, corresponding to lowest densities and largest lattice sizes used.}
\label{signplot}
\end{center}
\vspace{-6mm}
\end{figure}

A detailed description of our numerical setup is given in \cite{ourmain} and \cite{lat09}. We use the DDMC algorithm as introduced in \cite{burovski} with a few modifications which increase the efficiency by reducing autocorrelation effects that are present in the original setup. Also we generalise the algorithm to the spin-imbalanced case. If $\mu_\uparrow\neq\mu_\downarrow$, a sign problem arises, since the distribution function is no longer positive for all configurations. To deal with this problem we make use of the ``sign quenched method'', which is based on the ``phase quenched method'' known from lattice QCD \cite{phasequenched}. The idea is to write the thermal distribution as a product of its modulus and its sign, and to use the positive function given by the modulus as the new probability distribution. This implies a reweighting of each MC estimator, in this case simply a multiplication with the relative sign of the two matrix determinants. Additionally, each expectation value must be divided by the expectation value of the sign. If the latter is is close to zero, numerical errors will be very large, as it happens in QCD. However, for the unitary Fermi gas the sign remains very close to unity for small imbalances, as shown in Fig.~\ref{signplot}, so that sign quenching is applicable for imbalances up to approximately $\Delta\mu=0.2\varepsilon_F$. Our method can provide a useful tool to examine the trend of the critical temperature for small deviations from the balanced limit.

\vspace{-2mm}
\section{Results}
We obtained data at $25$ different values $(\mu_\uparrow,\mu_\downarrow)$, of which $8$ were at $\mu_\uparrow=\mu_\downarrow$. The lattice sizes varied between $4^3$ for the highest filling factor and $26^3$ for the lowest, so that the volume range in physical units was approximately constant. As discussed in \cite{burovski} for the balanced case, the dimensionless physical observables scale linearly with $\nu^{1/3}$ for sufficiently small $\nu$. With our data this behaviour is seen for $\nu^{1/3}\lessapprox0.75$. This condition was fulfilled for $23$ out of the $25$ points and in particular for $7$ out of the $8$ balanced points.

Since the chemical potential difference is less prone to numerical errors, we use $\Delta\mu/\varepsilon_F=|\mu_\uparrow-\mu_\downarrow|/\varepsilon_F$ instead of the relative density difference $\Delta\nu/\nu$ to quantify imbalance. For the values of imbalance considered in our study these two quantities are proportional to each other, with $\Delta\nu/\nu=0.122(2)\Delta\mu/\varepsilon_F$, see \cite{ourmain}. Every physical observable $X$ is a function of filling factor $\nu$~\nopagebreak and imbalance $h=\Delta\mu/\varepsilon_F$. We fit our data to a three dimensional surface, where the following assumptions are~\nopagebreak made for the form of the fitted function:
\begin{itemize}
\vspace{-2mm}
\item At fixed imbalance, $X$ is a linear function of $\nu^{1/3}$ with slope $\alpha^{(X)}(h)$: $X(\nu,h)=X(h)+\alpha^{(X)}(h)\nu^{1/3}$.
This is a generalisation of the relation valid in the balanced case.
\vspace{-2mm}
\item $X(h)$ and $\alpha^{(X)}(h)$ viewed as functions of the imbalance $h$ can be Taylor expanded.
\vspace{-2mm}
\item Due to symmetry in $h$ all odd powers in the expansions of $X(h)$ and $\alpha^{(X)}(h)$ have to vanish.
\vspace{-2mm}
\end{itemize}
If we expand $X(h)$ and $\alpha^{(X)}(h)$ to leading order in $h$ the fitted function becomes
\begin{equation}
X(\nu,h)=X_0+X_2h^2+(\alpha_0^{(X)}+\alpha_2^{(X)}h^2)\nu^{1/3}.
\label{imbquad}
\end{equation}
In the following we will present our fit results for several physical quantities. The results from sections \ref{subsectc}, \ref{subsecenperpart} and \ref{subsecchempot} were presented in more detail in \cite{ourmain}.

\vspace{-1mm}
\subsection{Critical Temperature}
\label{subsectc}
\begin{figure}
\vspace{-4mm}
\begin{minipage}[t]{.46\linewidth}
\includegraphics[width=\linewidth]{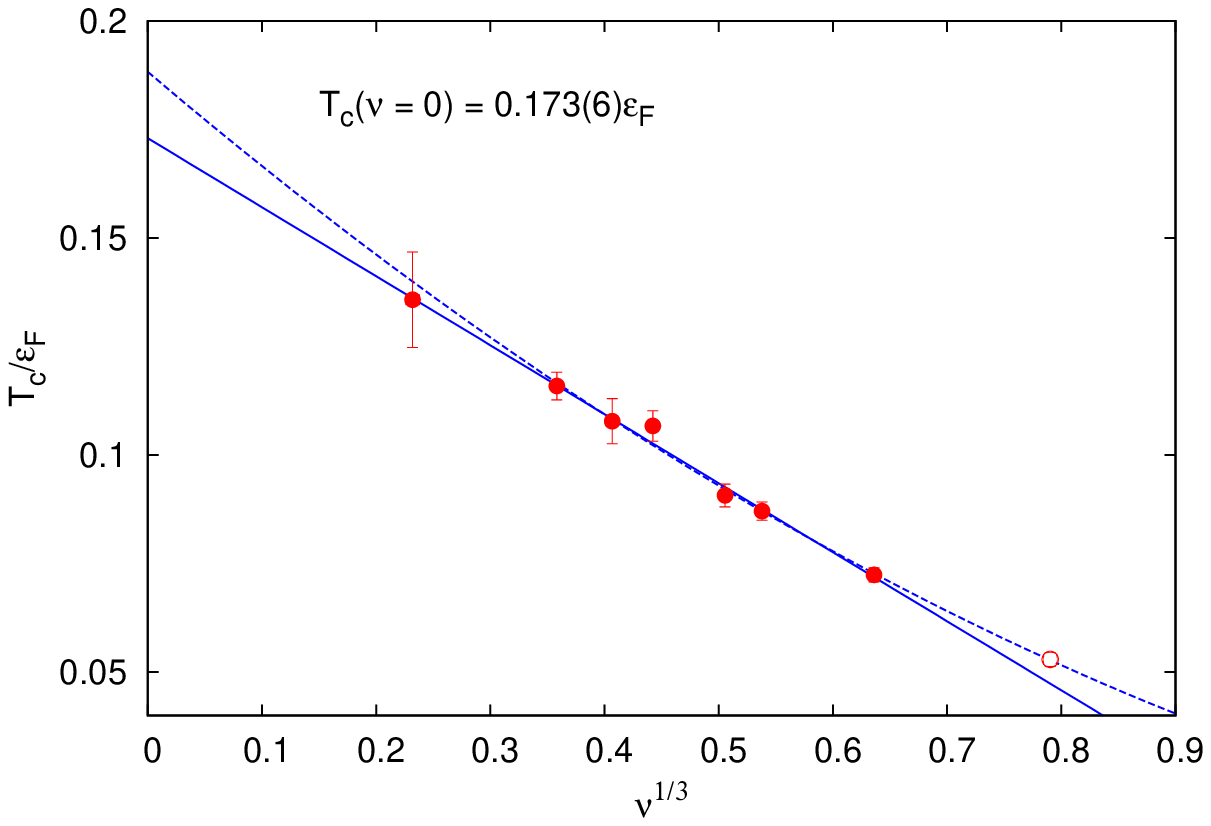}
\caption{$T_c$ versus $\nu^{1/3}$ for different values of the chemical potential at $\mu_\uparrow=\mu_\downarrow$. The solid line is the linear extrapolation of the $7$ data points with $\nu^{1/3}\lessapprox0.75$ (filled circles). The dashed line corresponds to a quadratic fit through all data points.}
\label{balancedtc}
\end{minipage}
\hfill
\begin{minipage}[t]{.46\linewidth}
\centerline{\includegraphics[width=\linewidth]{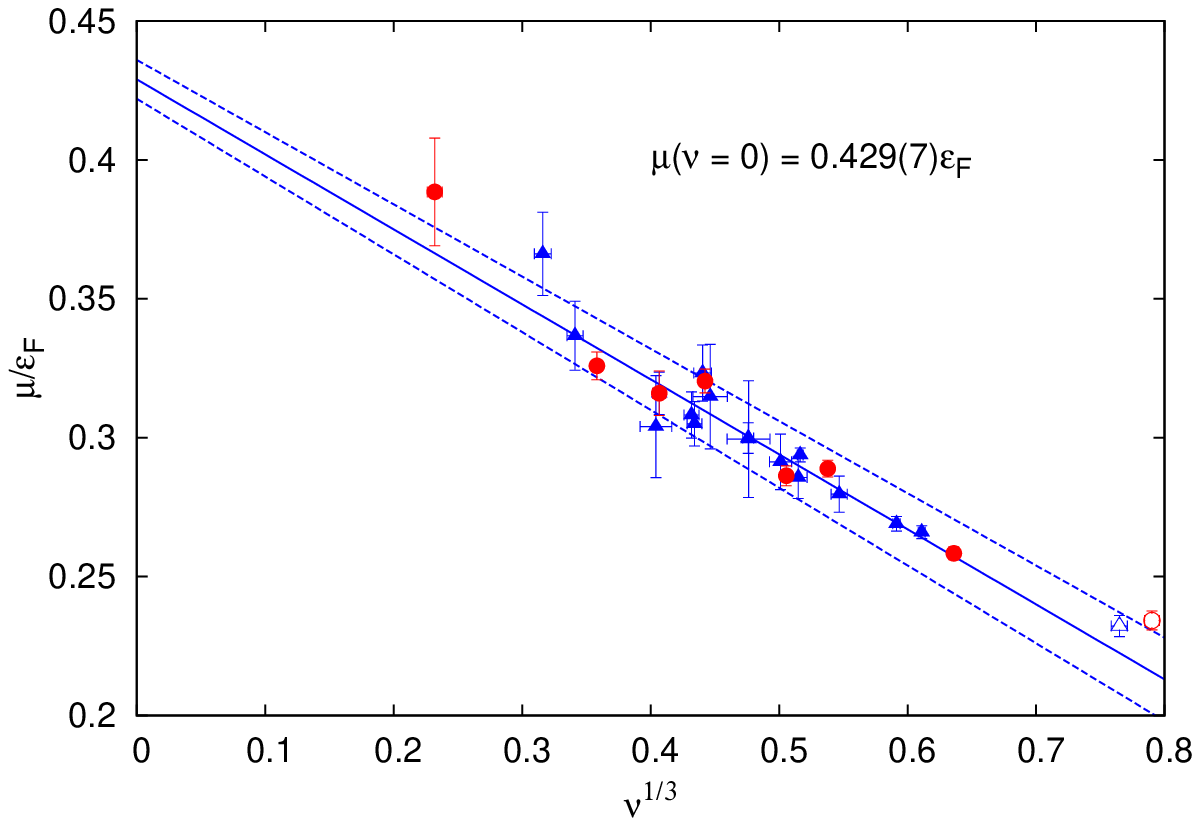}}
\caption{Projection of the data for the average chemical potential onto the $(\nu^{1/3}$-$\mu)$ plane. Red circles denote the balanced data and blue triangles data at non-zero imbalance. The solid line is the constant fit; dashed lines indicate uncertainty.}
\label{chempot}
\end{minipage}
\end{figure}

Fitting a line through the points with $\mu_\uparrow=\mu_\downarrow$ results in $T_c/\varepsilon_F=0.173(6)-0.16(1)\nu^{1/3}$ with $\chi^2/$d.o.f $=0.39$, as shown in Fig.~\ref{balancedtc}. For comparison we also fit a quadratic through all $8$ data points, resulting in a continuum value of $T_c/\varepsilon_F=0.188(15)$, which is in excellent agreement with the linear extrapolation. This confirms that sub-leading corrections proportional to $\nu^{2/3}$ can indeed be neglected for sufficiently small $\nu$.

Now we also include data with $\mu_\uparrow\neq\mu_\downarrow$. The best fit according to (\ref{imbquad}) yields $T_0=0.171(5)$, $\alpha_0^{(T)}=-0.154(9)$, $T_2=0.4\pm0.9$ and $\alpha_2^{(T)}=-0.7\pm1.9$ in units of $\varepsilon_F$, with $\chi^2/$d.o.f.$=0.43$. Note that the $T_2$ value corresponding to the minimal $\chi^2$ is positive, which is forbidden by physical arguments -- the critical temperature can only decrease with increasing imbalance. Since the $\chi^2$ function is very flat along the $T_2$ direction, forcing $T_2=0$ results in $\chi^2/$d.o.f.$=0.44$. From the error on $T_2$ we derive the lower bound $T_2>-0.5$. The best fit values for $T_0$ and $\alpha_0^{(T)}$ are in excellent agreement with the ones obtained from the fit of the balanced data only.

\begin{figure}
\vspace{-4mm}
\begin{center}
\includegraphics[width=0.48\textwidth]{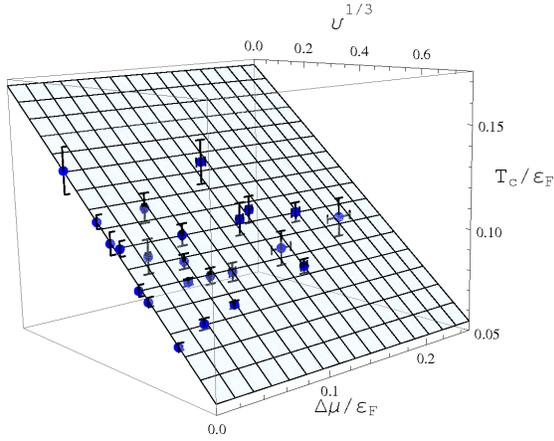}
\hfill
\includegraphics[width=0.48\textwidth]{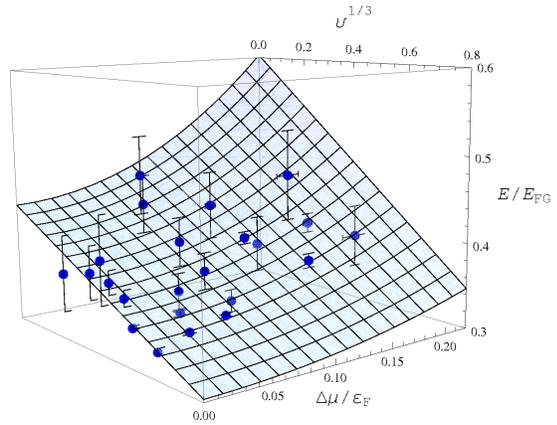}
\caption{$T_c/\varepsilon_F$ with the constant fit (left) and $E/E_{FG}$ with the quadratic fit (right) versus $\nu^{1/3}$ and $\Delta\mu/\varepsilon_F$.}
\label{3D}
\end{center}
\vspace{-5mm}
\end{figure}

By simplifying the fit model setting $\alpha_2^{(T)}=0$, the lower bound on $T_2$ can be tightened to $T_2>-0.04$. The other parameters $T_0$ and $\alpha_0^{(T)}$ agree with the results from the previous fit. This fit has $\chi^2/$d.o.f.$=0.41$ and is still consistent with $T_2=0$. For this reason we also perform a fit to constant $T_c(h)$ and $\alpha^{(T)}(h)$, and obtain $T_c(\nu,h)=0.1720(45)-0.156(8)\nu^{1/3}$ with $\chi^2/$d.o.f.$=0.41$. Again the result agrees with the previous fits. We also performed fits using the jackknife method and several robust fits and obtained consistent results. A three dimensional plot of the data together with the constant surface fit is presented in Fig.~\ref{3D} (left).

\vspace{-1mm}
\subsection{Energy per particle}
\label{subsecenperpart}
The total energy is composed of the kinetic energy $E_{\textnormal{kin}}=-\left\langle\sum_{\mathbf{x},\sigma}c^\dagger_{\mathbf{x}\sigma}\nabla^2c_{\mathbf{x}\sigma}\right\rangle$ and the interaction energy $E_{\textnormal{int}}=\langle H_1\rangle$. For the explicit MC estimators see \cite{ourmain}. Our results are obtained at $T_c$, but the temperature dependence of the energy per particle was found to be weak. Using only balanced data we obtain the continuum value $E/N\varepsilon_F=0.276(14)$, or $E/E_{FG}=0.46(2)$, in units of the ground state energy of the free gas $E_{FG}=(3/5)N\varepsilon_F$. The goodness of fit is $\chi^2/$d.o.f. $=2.1$. Since with increasing imbalance interactions become suppressed, the absolute value of the interaction energy must decrease. This in turn means an increase of the total energy, since the interaction energy is negative. As we did for the critical temperature we fit the energy in units of $E_{FG}$ to the function (\ref{imbquad}) and obtain the best fit parameters $E_0=0.440(15)$, $\alpha^{(E)}_0=-0.17(3)$, $E_2=3.4\pm2.2$ and $\alpha^{(E)}_2=-3.1\pm4.5$, with $\chi^2/$d.o.f.$=2.8$. These results are consistent with the balanced fit. Forcing $\alpha^{(E)}_2=0$ yields a best fit result of $E(\nu,h)=0.444(13)+1.9(3)h^2-0.18(2)\nu^{1/3}$ with $\chi^2/$d.o.f.$=2.7$, which agrees with the previous result. For a plot of the data see Fig.~\ref{3D} (right).

\vspace{-1mm}
\subsection{Chemical potential}
\label{subsecchempot}
For the chemical potential at $T_c$ we obtain the continuum value $\mu/\varepsilon_F=0.429(9)$ with $\chi^2/$d.o.f. $=2.8$ using only balanced data. A similar analysis can be performed for the average chemical potential $\mu/\varepsilon_F=|\mu_\uparrow+\mu_\downarrow|/2\varepsilon_F$ in presence of an imbalance. Since this quantity is not expected to depend on the imbalance we fit our data to a constant function and obtain $\mu(\nu,h)=0.429(7)-0.27(1)\nu^{1/3}$ in units of $\varepsilon_F$ with $\chi^2/$d.o.f.$=1.1$. This is in very good agreement with our balanced result. A plot of the data and the fit are shown in Fig.~\ref{chempot}.

\vspace{-1mm}
\subsection{Contact density}
Another important quantity is the contact density, which can be interpreted as a measure of the local pair density \cite{braaten}. The contact plays an important role for several universal relations derived by Tan \cite{tan}. The definition of the contact is $C=m^2g_0E_{\textnormal{int}}$, where $g_0$ is the physical coupling constant \cite{braaten, wernercastin}, and it is related to the contact density $\mathcal{C}$ via $C=\int\mathcal{C}(\mathbf{r})d^3r$, or for homogeneous systems simply $C=\mathcal{C}V$. The dimensionless quantity $\mathcal{C}/\varepsilon_F^2$ can be expressed as
\begin{equation}
\mathcal{C}/\varepsilon_F^2=(UE_{\textnormal{int}})/(4L^3\varepsilon_F^2).
\end{equation}
\begin{figure}
\vspace{-5mm}
\begin{center}
\includegraphics[width=0.51\textwidth]{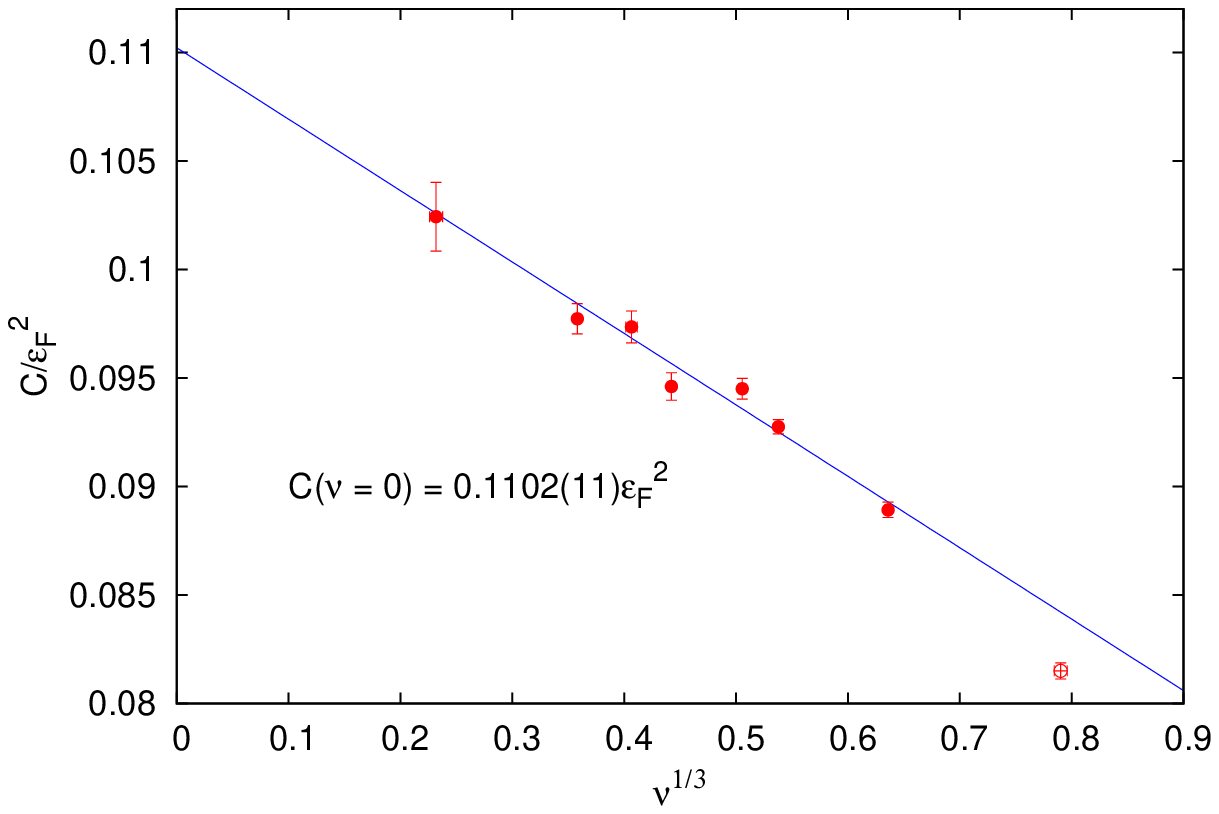}
\hfill
\includegraphics[width=0.42\textwidth]{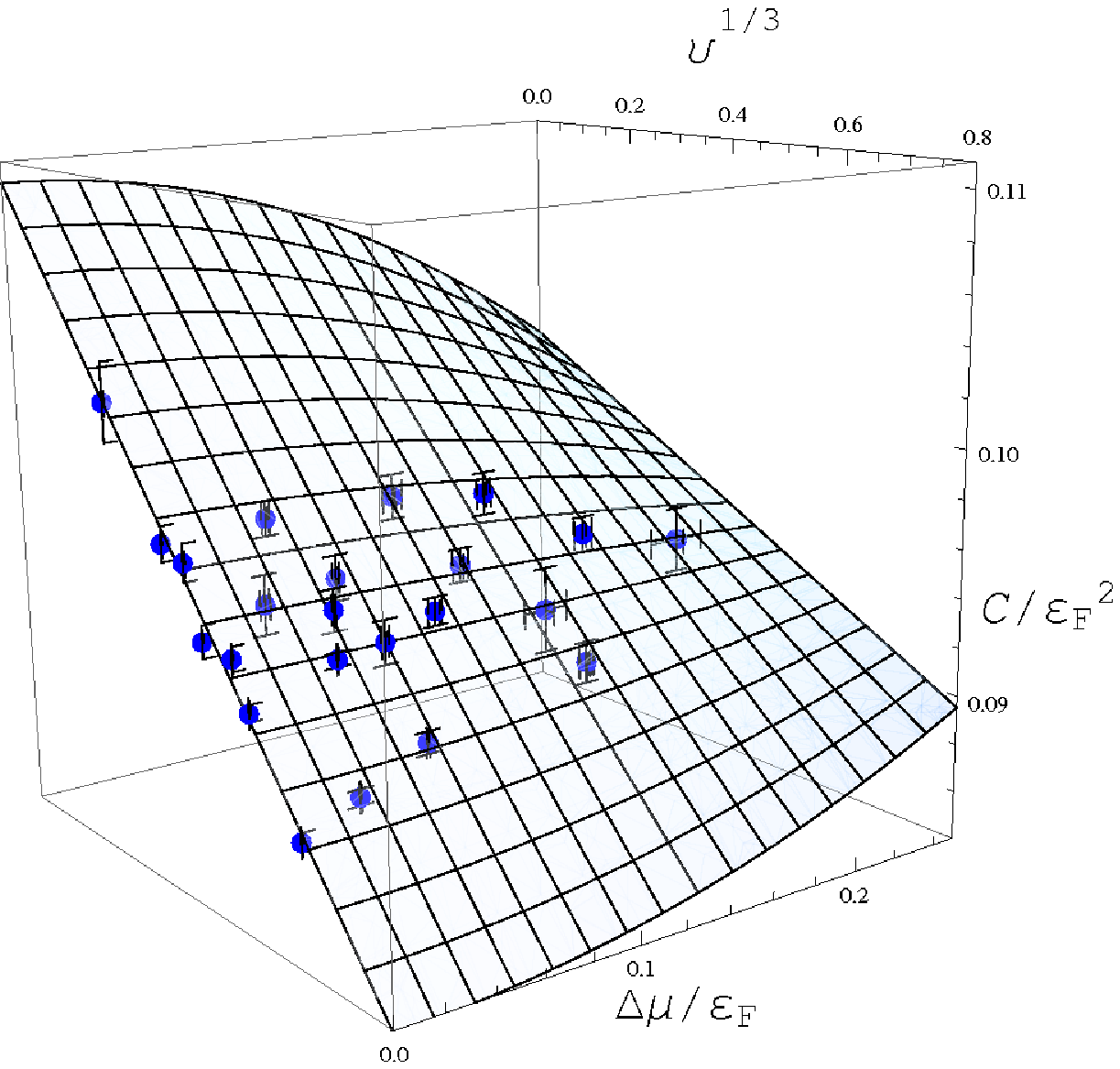}
\caption{Left: The contact density versus $\nu^{1/3}$ for balanced data together with the linear fit. Right: The contact density versus filling factor and imbalance. The surface corresponds to the quadratic fit.}
\label{contactfigure}
\end{center}
\vspace{-5mm}
\end{figure}
Using only balanced data the best fit is $\mathcal{C}/\varepsilon_F^2=0.1102(11)-0.033(2)\nu^{1/3}$ with $\chi^2/$d.o.f. $=1.8$. In the presence of an imbalance we expect the interaction energy and hence the contact density to decrease. A three dimensional fit of the data to (\ref{imbquad}) yields $C_0=0.1101(9)$, $\alpha^{(C)}_0=-0.033(2)$, $C_2=-0.15(16)$ and $\alpha^{(C)}_2=-0.29(36)$, with $\chi^2/$d.o.f.$=1.5$. This is consistent with the balanced fit. The parameter $C_2$ is negative as required. Forcing $\alpha^{(C)}_2=0$ yields $C_0=0.1099(8)$, $\alpha^{(C)}_0=-0.0322(15)$ and $C_2=-0.01(2)$, with $\chi^2/$d.o.f.$=1.5$. Fitting to a constant function yields $\mathcal{C}(\nu,h)=0.1097(8)-0.0320(14)\nu^{1/3}$ with $\chi^2/$d.o.f.$=1.4$. Figure \ref{contactfigure} summarises our results.

\vspace{-1mm}
\subsection{Comparison with the literature}
Our final result for $T_c$ using both balanced and imbalanced data, $T_c/\varepsilon_F=0.171(5)$, is significantly higher than the previous result from \cite{burovski}, where $T_c/\varepsilon_F=0.152(7)$. In \cite{burovskinew} the result of \cite{burovski} was found to agree with a continuous space-time DDMC method. The authors of \cite{bulgac1} found an upper bound of $T_c/\varepsilon_F\lessapprox0.15(1)$. They used an auxiliary field Monte Carlo approach and extracted $T_c$ using the same approximation as \cite{burovski} and \cite{burovskinew}, which might explain the discrepancy between our results. Through extrapolating Monte Carlo results of low-density neutron matter, the authors of \cite{abeTc} found $T_c/\varepsilon_F=0.189(12)$, which agrees with our result. There are also results obtained with the Restricted Path Integral Monte Carlo method \cite{rpimc}, $T_c/\varepsilon_F\approx0.245$, and an upper bound of $T_c/\varepsilon_F<0.14$ obtained with a hybrid Monte Carlo method \cite{lee}. Results from an $\epsilon$-expansion are also available \cite{epsexp}. For comparison, the critical temperature in the BEC limit is $T_{\textnormal{BEC}}=0.218\varepsilon_F$.

Our result for the energy per particle $E/E_{FG}=0.440(15)$ shows excellent agreement with the value $E/E_{FG}=0.45(1)$ at $T_c$ quoted in \cite{bulgac1}. The value quoted in \cite{burovski} is $E/N\varepsilon_F=0.31(1)$, which roughly corresponds to $E/E_{FG}=0.52(2)$. Our result for the chemical potential $\mu/\varepsilon_F=0.429(7)$ differs from $\mu/\varepsilon_F=0.493(14)$ quoted in \cite{burovski}, but is consistent with the value $\mu/\varepsilon_F=0.43(1)$ quoted in \cite{bulgac1}. Some theoretical predictions for the contact density are also available, but to our knowledge only at temperatures much lower than $T_c$ (see \cite{braaten} and references therein).

There are several recent experimental studies of the homogeneous unitary Fermi gas. The direct measurement presented in \cite{salomon} is $T_c/\varepsilon_F=0.157(15)$, which agrees well with our result, and $\mu/\varepsilon_F=0.49(2)$ at $T_c$, which differs from our value. The values from \cite{horikoshi}, $T_c/\varepsilon_F=0.17(1)$ and $\mu/\varepsilon_F=0.43(1)$ at $T_c$, show excellent agreement with our results. Their result for the energy per particle $E/N\varepsilon_F=0.34(2)$ at $T_c$ is higher than our value. In another experimental work \cite{MITexp} an estimate for $T_c$ at zero imbalance is extrapolated from data at higher values of imbalance. An experimental value for the contact density of the homogeneous unitary Fermi gas at zero temperature, $\mathcal{C}/\varepsilon_F^2=0.1184(64)$, is presented in \cite{salomoncontact}.

\vspace{-1mm}
\section{Outlook and Acknowledgements}
We have presented a Monte Carlo calculation of several thermodynamic observables of the unitary Fermi gas with equal and unequal chemical potentials in the two spin components. The improved DDMC algorithm with sign quenching also offers the intriguing possibility to explore the case of unequal masses of the two species. If $m_\uparrow\neq m_\downarrow$ the dispersion relations are different for the two components and the mass ratio enters as a new parameter. As in the spin-imbalanced case the two matrix determinants no longer need to be identical, so that sign quenching is required. The mass ratio is expected to influence the behaviour of the system significantly, so that exploring the phase diagram promises many new interesting insights.

This work used resources provided by the Cambridge High Performance Computing Facility. OG is supported by the German Academic Exchange Service (DAAD), the EPSRC and the CET.

\end{document}